\documentclass{article}
\usepackage{spconf,amsmath,graphicx}
\usepackage{siunitx}

\newcommand{\crnotice}{\copyright 2023 IEEE. Personal use of this material is permitted. Permission from IEEE must be obtained for all other uses, in any current or future media, including reprinting/republishing this material for advertising or promotional purposes, creating new collective works for resale or redistribution to servers or lists, or reuse of any copyrighted component of this work in other works.}
\newcommand{\pccy}{{\begin{tikzpicture} \node[text width=0.8\paperwidth, align=justify, anchor=south] {\crnotice}; \end{tikzpicture}}}
\usepackage{background}

\backgroundsetup{contents=\pccy, scale=1, angle=0, position={0.385\paperwidth, 15mm}, color=black, anchor=below}
\title{Relating EEG recordings to speech using envelope tracking and the speech-FFR}
\name{{Mike Thornton\textsuperscript{1}, Danilo Mandic\textsuperscript{1}, Tobias Reichenbach\textsuperscript{2}}\thanks{Mike Thornton is supported by UK Research and Innovation. [UKRI Centre for Doctoral Training in AI for Healthcare grant number EP/S023283/1].}}
\address{\textsuperscript{1}Imperial College London, \textsuperscript{2}Friedrich-Alexander-Universit{\"a}t Erlangen-N{\"u}rnberg}
\begin{document}
\ninept
\maketitle

\begin{abstract}
During speech perception, a listener’s electroencephalogram (EEG) reflects acoustic-level processing as well as higher-level cognitive factors such as speech comprehension and attention. However, decoding speech from EEG recordings is challenging due to the low signal-to-noise ratios of EEG signals. We report on an approach developed for the ICASSP 2023 ‘Auditory EEG Decoding’ Signal Processing Grand Challenge. A simple ensembling method is shown to considerably improve upon the baseline decoder performance. Even higher classification rates are achieved by jointly decoding the speech-evoked frequency-following response and responses to the temporal envelope of speech, as well as by fine-tuning the decoders to individual subjects. Our results could have applications in the diagnosis of hearing disorders or in cognitively steered hearing aids.
\end{abstract}
\section{Introduction}
\label{sec:intro}

When people listen to speech, their brainwaves synchronise with acoustic features such as the speech envelope. The degree of this neural tracking reflects cognitive factors such as attention to speech, speech comprehension, and intelligibility~\cite{Etard5750, Accou2021}. Accurately decoding speech from electroencephalography (EEG) is, however, a challenging task owing to the low signal-to-noise ratios of EEG recordings and the limited availability of  EEG data recorded during speech perception.

Neural speech tracking in a particular subject is best assessed when EEG responses to speech from that particular subject are available and can be used to train a subject-specific decoder.   Decoding EEG responses to speech from unseen subjects is a harder task, since EEG signals vary greatly between individuals. The ICASSP 2023 ‘Auditory EEG Decoding’ Signal Processing Grand Challenge involves both types of decoding. Its task is to  develop population match-mismatch decoders: given a temporal segment of EEG data and two candidate speech segments, the decoder should predict which of the speech segments corresponds to the EEG signal. This should be done both for EEG recordings from subjects that have been included in the training set, and for others whose EEG data has not been seen in the training stage.

Here we describe the development of our decoders which placed first in this match-mismatch task. Besides the speech envelope, we relate a second acoustic feature to the EEG recordings. This feature is the temporal fine structure of the voiced parts of speech (which consist of a fundamental frequency and many higher harmonics). The electroencephalogram displays a strong response at the fundamental frequency of speech, termed the speech-frequency-following response or speech-FFR~\cite{Kraus}. The speech-FFR is driven both by the fundamental frequency itself as well as by higher harmonics, and can be decoded from EEG responses to continuous speech~\cite{Kraus,forte2017human,Etard,Kulasingham2020}.

\section{Materials and methods}
\label{sec:methods}

\textit{2.1. Dataset.} A large training dataset was provided by the ICASSP competition organisers, which consisted of EEG recordings from 71 subjects who listened to speech material~\cite{K3VSND_2023}. Decoders were evaluated against a heldout dataset comprising EEG from 70 subjects included in the training dataset, and 15 new unseen subjects.

We used two pre-processed versions of the dataset that contained the two speech features of interest together with the corresponding EEG signals. The first dataset consisted of speech envelopes and EEG recordings sampled at \SI{64}{\hertz}. The second dataset contained the envelope modulations of the higher harmonics of the fundamental frequency of the voiced parts of speech, together with EEG sampled at \SI{512}{\hertz}~\cite{Kulasingham2020}. Pre-processing and methodological details will be described in a forthcoming publication~\cite{Thorn2023}.\\[-5pt]

\noindent\textit{2.2. Baseline decoder.} The baseline decoder is a deep neural network (DNN) proposed by Accou~\textit{et~al.}~\cite{Accou2021}. This DNN brings the speech envelopes and EEG recordings into a space where the matched envelope and EEG segments are maximally similar. The output layer consists of a single Sigmoid neuron. The matched and mismatched envelopes are presented as an ordered pair, and the Sigmoid neuron predicts the probability that the first envelope is matched to the EEG. The decoder is trained with the binary crossentropy loss function and Adam optimizer without regularisation.\\[-5pt]

\noindent\textit{2.3. Baseline + speech-FFR decoder.} The baseline decoder relates the EEG recordings to the temporal envelope of speech only. We were also interested in relating the EEG signals to the temporal fine structure of speech, since this can improve classification performance~\cite{lirias3836010}. We retained the architecture of the baseline decoder, but swapped the speech envelopes for the high-frequency envelope-modulations feature~\cite{Kulasingham2020}. The Sigmoid outputs of this speech-FFR decoder and the baseline decoder were combined via linear discriminant analysis (LDA) to produce the final predicted label.\\[-5pt]

\noindent\textit{2.4. Decoder training and fine-tuning.} Training examples were presented to the decoders as temporal segments of \SI{3}{\second} in duration (the same duration was used for evaluation). Sources of randomness in the training procedure include the decoder initialisation, and the order in which training examples were presented. The effects of these were marginalised by averaging the Sigmoid outputs of several trained instances of the decoders. For the population decoders, the hop length between the onsets of the training examples was \SI{1}{\second}. When fine-tuning the population decoder to individual subjects, this was reduced to \SI{0.125}{\second}, and regularisation (batch normalisation and spatial dropout) was applied to the input layers.

\section{Results}
\label{sec:results}

\begin{figure}[h!]
    \centering
    \includegraphics[width=\linewidth]{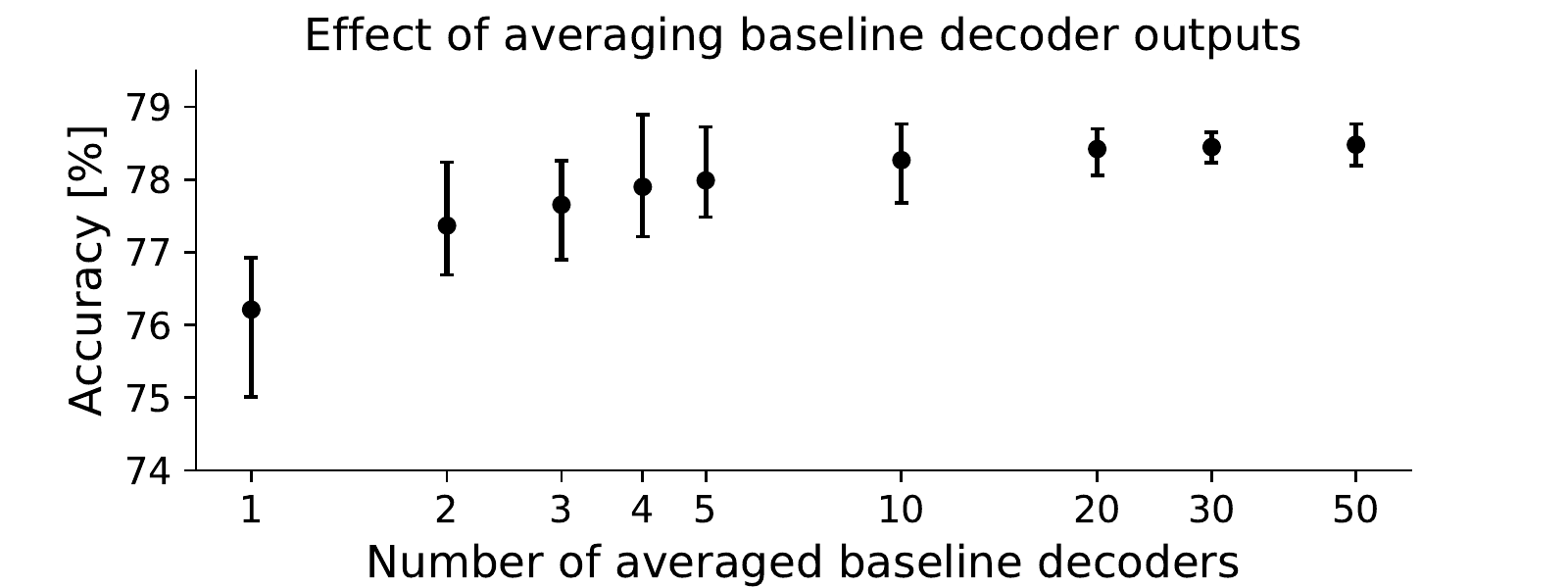}
    \caption{Classification accuracies against number of averaged decoders (log~scale). The bootstrapped mean and range of the accuracies are denoted by the dots and errorbars respectively.}
    \label{fig:avging}
\end{figure}

\textit{3.1. Averaging of decoder outputs.} We trained 100 instances of the baseline decoder. By averaging the Sigmoid outputs of the instances, the classification accuracy was improved (Figure~\ref{fig:avging}). Therefore, we formed two averaged population decoders: these used 50 instances of the baseline decoder, and 30 instances of the speech-FFR decoder, respectively. The averaged decoders were used in Section~3.2.

\begin{figure}[h!]
    \centering
    \includegraphics[width=0.49\linewidth, trim={1cm 0.5cm 2cm 0},clip]{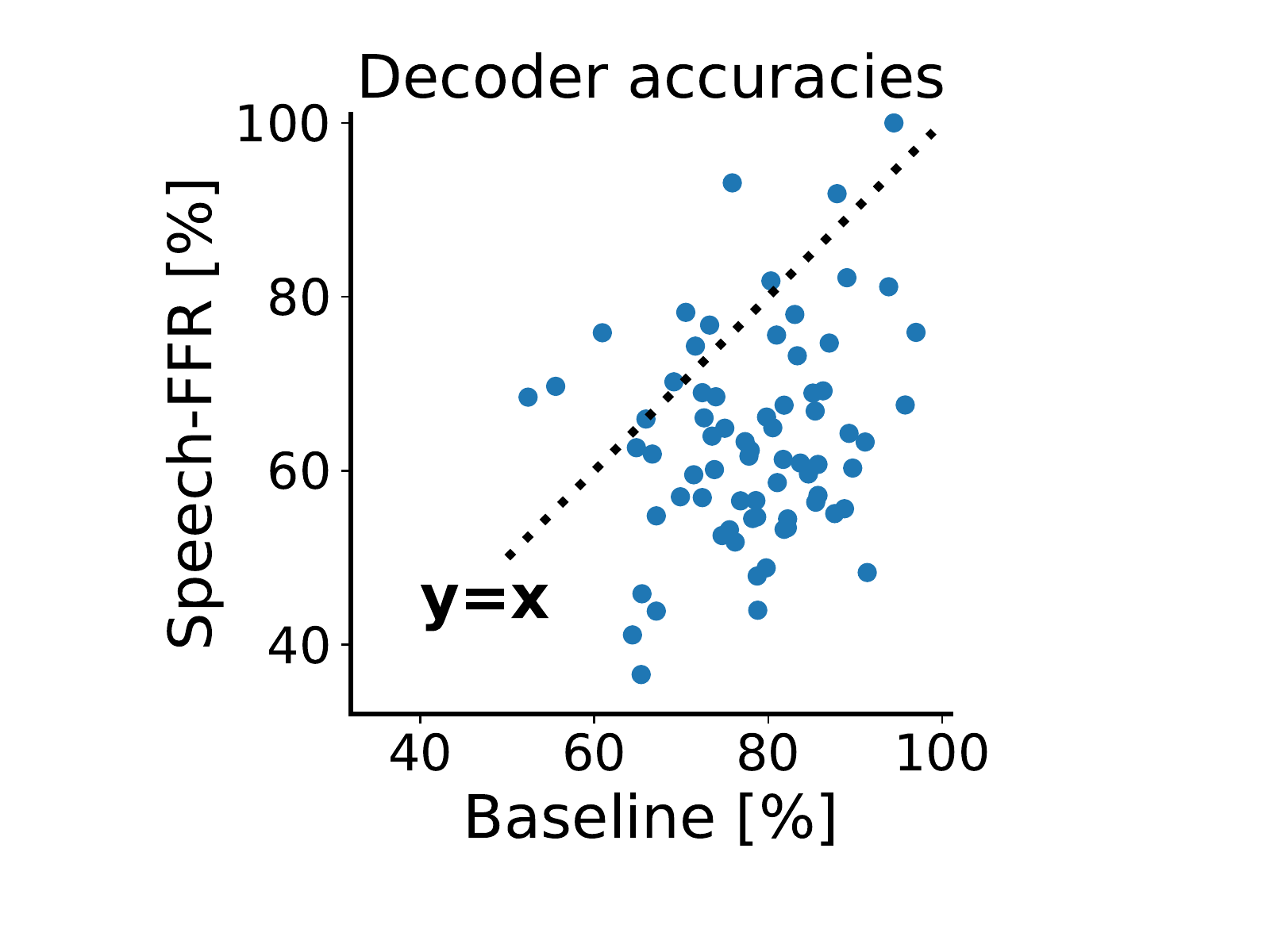}
    \includegraphics[width=0.49\linewidth, trim={1cm 0.5cm 2cm 0},clip]{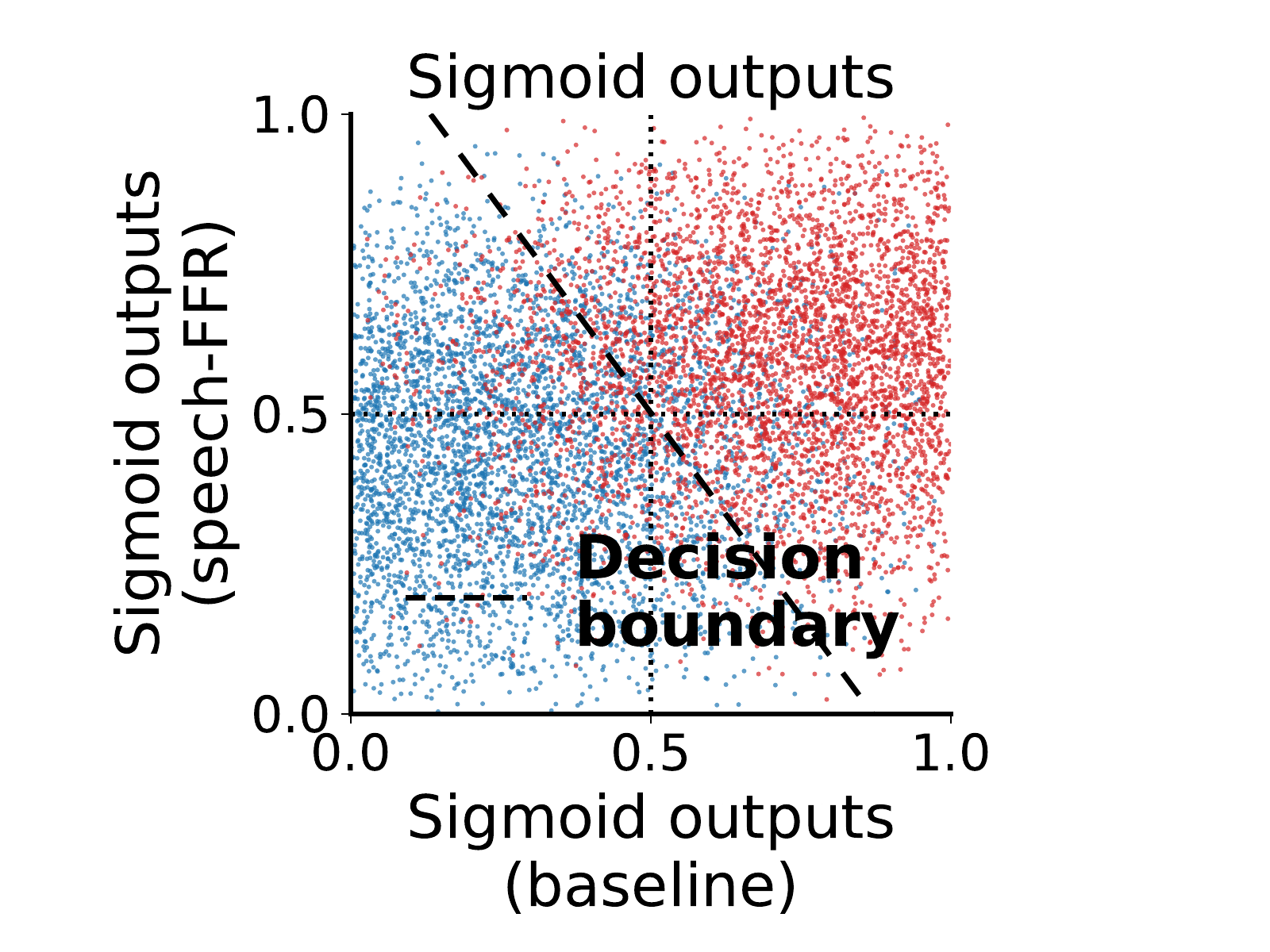}
    \caption{Comparison between the averaged baseline and speech-FFR decoders. (Left) The accuracies of the two decoders are shown for the 71 subjects. (Right) The Sigmoid outputs of the averaged decoders are shown. A red (blue) point indicates an EEG segment that matches the first (second) speech segment (recall that these are grouped as an ordered pair).  The LDA decision boundary is shown.}
    \label{fig:ffrVenv}
\end{figure}

\noindent \textit{3.2. Combining averaged baseline and speech-FFR decoders.} The baseline decoder generally achieves higher accuracies than the speech-FFR decoder (Figure~\ref{fig:ffrVenv},~left). The LDA classifier was fitted on an unseen portion of the training dataset, for which the correlation between the Sigmoid outputs of the decoders was moderate ($R=0.229$, Figure~\ref{fig:ffrVenv},~right). For the heldout dataset the correlation was similar ($R=0.206$), confirming that there was no severe overfitting or distributional shift. This composite decoder achieved an accuracy of 81.18\% on the heldout dataset for unseen subjects.

\begin{figure}
    \centering
    \includegraphics[width=\linewidth]{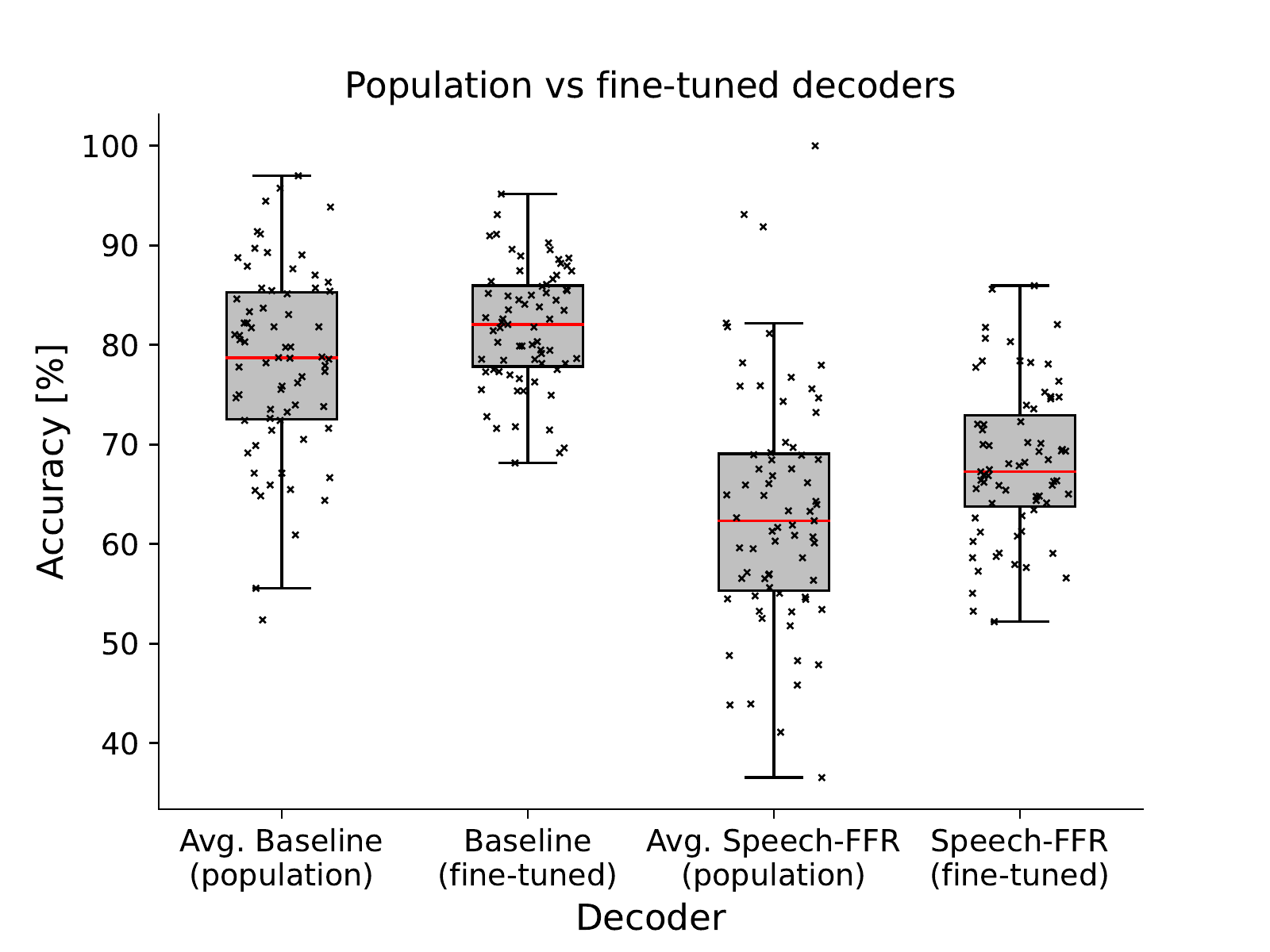}
    \caption{Effect of fine-tuning the decoders to individual subjects. Each datapoint represents the classification accuracy for a single subject.}
    \label{fig:finetuning}
\end{figure}

\noindent \textit{3.3. Decoder fine-tuning.} Figure~\ref{fig:finetuning} shows the effect of fine-tuning the decoders to individual subjects. For both the baseline decoders and the speech-FFR decoders, fine-tuning significantly improved the classification accuracy ($p<0.001$, signed-rank tests). The fine-tuned baseline decoders achieved an accuracy of 82.71\% on the heldout dataset for seen subjects.

\section{Conclusions}
\label{sec:discussion}
The performance of the baseline decoder could be improved by averaging the outputs of several trained decoder instances. Combining the averaged baseline and speech-FFR decoders enhanced the decoding accuracies  further. The best results were obtained by using versions of the baseline decoder which were fine-tuned to individual subjects where possible, and by using the composite decoder for unseen subjects. This approach won the match-mismatch task of the ICASSP 2023 Signal Processing Grand Challenge ‘Auditory EEG Decoding’. Future work will establish which aspects of the fine-tuning procedure led to such high classification accuracies.


\bibliographystyle{IEEEbib}
\bibliography{main}

\end{document}